\documentclass[preprint,12pt]{elsarticle}
\usepackage{amssymb}
\usepackage{amsmath}
\usepackage{caption}
\usepackage{subcaption}
\journal{Journal of Computational Physics}

\begin{document}

\begin{frontmatter}
\title{A volume of fluid method for simulating fluid/fluid interfaces in
contact with solid boundaries}
\author[]{Kyle Mahady, Shahriar Afkhami~\footnote{Corresponding author email
address: shahriar.afkhami@njit.edu}, and Lou Kondic}
\address{Department of Mathematical Sciences, New Jersey Institute of
  Technology, Newark, NJ
  07102-1982, USA}

\begin{abstract}
In this paper, we present a novel approach to model the fluid/solid interaction
forces in a direct solver of the Navier-Stokes equations based on the volume of
fluid interface tracking method. The key ingredient of the model is the
explicit inclusion of the fluid/solid interaction forces into the governing
equations. We show that the interaction forces lead to a partial wetting
condition and in particular to a natural definition of the equilibrium contact
angle.   We present two numerical methods to discretize the interaction forces
that enter the model; these two approaches differ in complexity and
convergence.  To validate the computational framework, we consider the
application of these models to simulate two-dimensional drops at equilibrium,
as well as drop spreading.  We demonstrate that the model, by including the
underlying physics, captures contact line dynamics for arbitrary contact
angles.  More generally, the approach permits novel means to study contact
lines, as well as a diverse range of phenomena that previously could not be
addressed in direct simulations.
\end{abstract}


\end{frontmatter}
\section{Introduction}

The interaction of fluid/fluid interfaces with solid boundaries is of
fundamental importance to a variety of wetting and dewetting phenomena.   In
this paper, we present a computational framework for the inclusion of a general
fluid/solid interaction, treated as a temporally and spatially dependent body
force,  in a direct solver of the Navier-Stokes equations.   This approach
allows for computing fluid wetting properties (such as equilibrium contact
angle) based on first principles, and without restriction to small contact
angles.   

Due to the complexities involved in modeling dynamics of fluids on solid
substrates, a significant amount of modeling and computational work
has been carried out using the long-wave (lubrication) approach.
Still, even within the long-wave approach, a difficulty arises when employing the 
commonly used no-slip boundary condition at the fluid/solid interface:
a non-integrable shear-stress singularity at the moving contact line.
Simulating dynamic contact lines therefore requires additional ingredients
for the model.
One option is to include fluid/solid interaction forces with conjoining-disjoining
terms which lead to a prewetted (often called `precursor') layer in
nominally `dry' regions.
This approach effectively removes the `true' contact line, consequently
avoiding the associated singularity~\cite{Schw98,dk_pof07,eggers2005}.
A second approach is to relax the no-slip condition and instead assume the
presence of slip at the fluid/solid interface. 
Both slip and disjoining
pressure approaches have been extensively used to model a variety of problems
including wetting, dewetting, film breakup, and many others (see
e.g.~\cite{oron_rmp97,cm_rmp09} for reviews).

While the approach based on the long-wave model has been very successful, it
does include limitations inherent in its formulation: in particular, the
restriction to small interfacial slopes (strictly speaking, the slopes much
less than unity), and therefore small contact angles.  We have shown in our
earlier work~\cite{mahady_13} 
that, depending on the choice of flow geometry, the
comparison between the solutions of the long-wave model and of the
Navier-Stokes equations may be better than expected; however, still for
slopes of $O(1)$, quantitative agreement disappears. 
Therefore, one would
like to be able to consider wetting/dewetting problems by working outside of
the long-wave limit, while still considering the most important physical
effects; such as fluid/solid interaction forces.
These fluid/solid interactions are known to be crucial in determining stability
properties of a fluid film; without their presence, a fluid film on a substrate is 
stable, since there are no forces in the model to destabilize it. 
In particular, for thin nanoscale films, fluid/solid interaction forces may be dominant.
We note that the approaches based on the Derjaguin approximation to
include the van der Waals or electrostatic interactions into the model in the
form of a local pressure contribution (disjoining pressure) acting on the
fluid/solid interface are derived under the assumption of a flat
film~\cite{Israelachvili,carey99,Derjaguin41,Churaev95}.  Therefore these
approaches cannot be trivially extended to the configurations involving large contact
angles.

In the context of direct simulations of the Navier-Stokes equations with free 
interfaces, a large variety of methods are used to track the evolution of the interface.
Lagrangian methods conform the computational grid to the
interface (e.g.~\cite{Baer2000,Sprittles12,Sprittles2013}).
Eulerian methods require a separate mechanism to track the interface
location; these include front tracking methods (e.g.~\cite{Unverdi1992}),
and interface capturing methods such as volume of fluid methods and 
level set methods.
The latter two methods easily treat topology changes, and with
recent developments have been shown to be effective for simulating surface
tension driven flows~\cite{Sussman99,Prost,FCDKSW2006,Popinet2009}.
A common feature of volume of fluid methods is that contact angles are imposed
geometrically, in that the angle at which the interface intersects the solid
substrate is specified as a boundary condition on the
interface~\cite{AB2008,AB2009,Sussman2012}.  Such approaches were used  to model 
the dynamics of non-wetting drops, that
could even detach from the substrate~\cite{afkhami-kondic-2013}, as well as
spreading drops~\cite{Sussman2012,Spelt2005,afkhami_jcp09}.  The van der Waals
interaction has been implemented previously in a volume of
fluid based solver for the liquid/liquid interaction of colliding
droplets~\cite{jiang2007}, but to our knowledge has not been considered for
flows involving wetting phenomena.

A variety of other computational methods have been considered in the context of
wetting/dewetting.  Here we mention phase-field methods that treat two fluids
with a diffuse interface by means of a smooth concentration function, which
typically satisfies the Cahn-Hilliard or Allen-Cahn equations, and is coupled
to the Navier-Stokes equations.  Jacqmin~\cite{Jacqmin1999} describes
a phase-field contact angle model that use a wall energy to determine the value
of the normal derivative of the concentration on a solid substrate. This model
has been used to study contact line dynamics~\cite{Jacqmin2000,Jacqmin2004},
and similar models have been considered in the investigation of the sharp
interface limit of the diffuse interface model~\cite{Yue2010,Sibley2013}.
Lattice-Boltzmann methods have also treated the contact angle with a wall
energy contribution~\cite{Briant2004,Lee2010}. 
These approaches have explained a variety of phenomena related to spreading of
fluids on solid substrates, but do not consider explicitly the stabilizing and
destabilizing forces between fluid and solid, as has been done via disjoining
pressure within the context of the long-wave model. The
liquid/solid interaction is naturally included in molecular dynamics (MD)
simulations~\cite{Qian2006,Nguyen2012,fuentes_pre11} that typically consider
Lennard Jones potential between fluid and solid particles. However, MD
simulations are, in general, computationally expensive, even when simulating
nanoscale systems.  One would like to be able to include liquid-solid
interaction within the framework of a continuum model. 

Here, we present a novel approach, based on a volume of fluid formulation, 
which includes the fluid/solid interaction forces
into the governing Navier-Stokes equations, without limitations inherent in the
long-wave model. 
This inclusion allows for arbitrary contact angles to be incorporated based on
modeling the underlying physics, in contrast to conventional volume of fluid
methods. 
The presented approach also
leads to the regularization of the viscous stress since the fluid film
thickness never becomes zero. Furthermore, our framework can account for
additional physical effects, such as instability and breakup of thin fluid
films, that would not be described if fluid/solid interaction forces were not
explicitly included.  We note here that while film rupture can also occur in
phase-field based approaches (as in~\cite{Jacqmin1999}), this seems to be due to  
the presence of a rather thick interface, and not due to the explicit inclusion
of destabilizing liquid/solid interaction forces.   

In the present paper we focus on formulating and discretizing the model, and on
discussing issues related to convergence and accuracy. To validate our proposed
numerical scheme, we consider two representative examples, involving relaxation
and spreading of sessile drops with various contact angles on a substrate.
These benchmark cases permit comparison of our results with well established
analytical solutions for a particular flow regime.  The application of the
method to the study of thin film stability including dewetting will be
considered in the sequel~\cite{MahadyvdW2015}.   

The rest of this paper is organized as follows.  We describe the details of the
fluid/solid interaction in Sec.~\ref{sec:model}.  In Sec.~\ref{sec:methods}, we
describe two finite-volume methods for the discretization of the considered
fluid/solid interaction forces.  The presentation in these two sections applies
to any generic fluids.  In Sec.~\ref{sec:results} we present a comparison of
the two considered discretization methods for equilibrium and spreading drops,
for a particular choice of material parameters.  In Sec.~\ref{sec:conc}, we
give an overview and future outlook.   

\section{Model}\label{sec:model}
Consider a perfectly flat solid substrate covered by two immiscible fluids.  For clarity,
we will refer to these fluids as the liquid phase (subscript $l$), and the vapor phase
(subscript $v$), although the present formulation applies to any two fluids.  Assume that
gravity can be neglected, and also ignore any phase-change effects, such as evaporation
and condensation.  There are three relevant interfacial energies: the liquid/solid,
$\gamma_{ls}$, the vapor/solid, $\gamma_{vs}$, and the liquid/vapor,  $\gamma$, energies.
The contact angle is commonly defined as the angle between the tangent plane of the
interface between the liquid and vapor phases and the solid substrate at the point where
the interface meets the surface.  At equilibrium, the contact angle, $\theta_{eq}$, and
the surface energies are related by Young's Equation~\cite{Young}:
\begin{equation}\label{eq:youngs}
	\gamma_{vs} = \gamma_{ls} + \gamma\cos\theta_{eq}
\end{equation}
If there is a nonzero contact angle at equilibrium, the liquid partially wets the solid
surface; if the equilibrium configuration is a flat layer covering the whole substrate,
then it fully wets the solid surface. It is also possible for the liquid to be
non-wetting, where the liquid beads up into a sphere on the surface.  The wetting behavior
of the system can be characterized by the equilibrium spreading coefficient, defined by:
\begin{equation}\label{eq:spreading_coeff}
	S_{eq} = \gamma_{vs} - (\gamma_{ls} + \gamma)
\end{equation}
which expresses the difference in energy per unit area between a surface with no liquid,
and one with a layer of liquid (what we call `dry' and `wet' states, respectively).
Wetting is determined by the sign of $S_{eq}$; partial wetting occurs for $S_{eq}<0$, and
complete wetting for $S_{eq}=0$~\cite{bonn_rmp2009}.

The above characterization of the contact angle is straightforward for static
configurations and at macroscopic length scales. If these assumptions are not satisfied,
definitions of contact angles become more complex. In the literature, a distinction is
made between the apparent contact angle, $\theta_{ap}$, and the microscopic contact angle,
$\theta_m$, distinguished by the distance from the contact line at which the measurement
is made~\cite{Dussan79}.  The contact angle resulting from measuring on macroscopic length
scales is $\theta_{ap}$, while $\theta_m$ is measured at short length scales which are
still long enough so that the continuum limit is appropriate~\cite{bonn_rmp2009,voinov1976,Shikhmurzaev}.
The microscopic contact angle, $\theta_m$, is often identified with
$\theta_{eq}$, which is commonly used in the derivation of spreading laws, such as the
classical Cox-Voinov law~\cite{voinov1976}.     It should be noted that the details of the
fluid behavior on nano scales in the vicinity of the fluid fronts and associated contact
lines are far from being completely understood~\cite{bonn_rmp2009}, and the way in which
the contact angle arises at small scales is a subject of ongoing
research~\cite{snoeijer2008}.

The surface energies entering Eq.~\eqref{eq:youngs} and Eq.~\eqref{eq:spreading_coeff}
arise due to the van der Waals interaction between the different phases that are relevant
on short length scales.  Three kinds of van der Waals interactions are usually considered:
interactions between polar molecules, interaction of molecules that have an induced
polarization, and the dispersion forces~\cite{Israelachvili}.  The dispersion interaction
is relevant for all molecules, and will be the only one that we consider.  A common model
for approximating the dispersion interaction of two particles, of phases $i$ and $j$,
centered at at $\textbf{x}_0$ and $\textbf{x}_1$ is the 12-6 Lennard Jones
potential~\cite{Israelachvili}:
\begin{equation}\label{eq:126LJ}
	\phi_{\text{LJ}}(r) = 4\epsilon_{ij}\left( \left(\frac{\sigma}{r}\right)^{12} - \left(\frac{\sigma}{r}\right)^6 \right)
\end{equation}
This potential has a minimum, $\epsilon_{ij}$, at $r=2^{1/6}\sigma$, and for simplicity we
assume that $\sigma$ is a constant for any two interacting particles.  The center distance
between the two particles is given by:
$$
	r = \sqrt{(x_0-x_1)^2 + (y_0-y_1)^2 + (z_0-z_1)^2}
$$
The powers $12$ and $6$ in Eq.~\eqref{eq:126LJ} correspond to short range repulsion and
long range attraction, respectively.  For our purposes, we generalize this formulation and
use the following form:
\begin{equation}\label{eq:potential}
	\phi_{ij}(r) = K^\ast_{ij}\left( \left(\frac{\sigma}{r}\right)^p - \left(\frac{\sigma}{r}\right)^q \right)
\end{equation}
Here $K^\ast_{ij}$  is the scale of the potential well, having units of energy per
particle pair interaction. In general, we require only that $p>q$ are integers, such that
$q>3$, for reasons that will become clear shortly.

Our model considers two fluid phases, a liquid phase and a vapor phase occupying the
region $y>0$, interacting with a flat, half infinite, solid substrate (subscript $s$) in
the region $y<0$.  Consider now a particle located at $\textbf{x}_0 = (x_0, y_0, z_0)$ of
phase $i$ (where $i$ is either $l$ or $v$).  The interaction energy between this particle
and the substrate per unit volume of the substrate is:
\begin{equation}\label{eq:int_per_vol_subst}
	\psi_{is}(r) = n_s \phi_{is}(r)
\end{equation}
where $n_s$ is the particle density in the substrate.  

We derive the force per unit volume following a similar procedure outlined
in~\cite{jiang2007}.  Integrating Eq.~\eqref{eq:int_per_vol_subst} over the region $y<0$
yields the following total interaction of a particle in phase $i$ with the substrate:
\begin{multline*}
	\int_{-\infty}^0 \int_{\infty}^{\infty} \int_{-\infty} ^{\infty} \psi_{is}(r) dx dz dy = 2\pi n_s K^\ast_{is}\sigma^3\left[\frac{1}{(2-p)(3-p)}\left(\frac{\sigma}{y_0}\right)^{p-3} \right.\\ 
		\left.- \frac{1}{(2-q)(3-q)}\left(\frac{\sigma}{y_0}\right)^{q-3}  \right]
\end{multline*}
Although the van der Waals interaction is not strictly additive, the effects due to
non-additivity are usually weak~\cite{Israelachvili}, and we ignore them for simplicity. 

Multiplying by $n_i$, the particle density in phase $i$, we obtain the van der Waals
interaction per unit volume of phase $i$: 
\begin{multline}\label{eq:int_per_vol}
	\Phi_{is}(y_0) = 2\pi n_i n_s K^\ast_{is}\sigma^3\left[\frac{1}{(2-p)(3-p)}\left(\frac{\sigma}{y_0}\right)^{p-3} \right. \\ 
		\left.-  \frac{1}{(2-q)(3-q)}\left(\frac{\sigma}{y_0}\right)^{q-3}\right]
\end{multline}
We introduce the following parameters:
\begin{equation}\label{eq:strength}
	\mathcal{K}_{is} = 2\pi n_i n_s K^\ast_{is}\sigma^3 \left(\frac{\left[(p-2)(p-3)\right]^{q-3}}{\left[(q-2)(q-3)\right]^{p-3}}\right)^{\frac{1}{p-q}}
\end{equation}
\begin{equation}\label{eq:equifilm}
	h^\ast = \left[\frac{(q-2)(q-3)}{(p-2)(p-3)}\right]^\frac{1}{p-q}\sigma
\end{equation}
\begin{eqnarray*}
	m = p-3	&& n=q-3
\end {eqnarray*}
This simplifies Eq.~\eqref{eq:int_per_vol} to:
\begin{equation}\label{eq:interaction}
	\Phi_{is}(y) = \mathcal{K}_{is}\left[\left(\frac{h^\ast}{y}\right)^m - \left(\frac{h^\ast}{y}\right)^n\right]
\end{equation}
The quantity $h^\ast$ is referred to as the `equilibrium film thickness' and will be
considered in more detail below.  Note that Eq.~\eqref{eq:interaction} has an identical
form as Eq.~\eqref{eq:potential}.

The force per unit volume on phase $i$ that results from the potential is computed by
taking the gradient of Eq.~\eqref{eq:interaction}:
\begin{equation}\label{eq:force}
	\textbf{F}_{is}(y) = -\nabla \Phi_{is}(y) = \frac{\mathcal{K}_{is}}{h^\ast}\left[m\left( \frac{h^\ast}{y} \right)^{m+1} - n\left(\frac{h^\ast}{y} \right)^{n+1} \right]\hat{y}
\end{equation}
Here $\hat{y}$ refers to the unit vector $(0,1,0)$.

We proceed to derive an expression for $\theta_{eq}$ in terms of $\mathcal{K}_{ls}$ and
$\mathcal{K}_{vs}$ in Eq.~\eqref{eq:interaction} according to the microscopic arguments
outlined in~\cite{carey99}.  Combining Eqs.~\eqref{eq:youngs} and
~\eqref{eq:spreading_coeff}, we obtain the following expression for the equilibrium
spreading coefficient:
\begin{equation}\label{eq:spreading_II}
	S_{eq} = \gamma(\cos\theta_{eq} - 1)
\end{equation}
where $S_{eq}$ is the difference in the energy per unit area between a dry system and a
wet system.  The total energy required to remove a liquid layer originally occupying
$\delta_0 < y < \infty$ and replace it with a layer of the vapor phase is given by:
\begin{equation}\label{eq:energy_diff}
	\Delta E= \int_{\delta_0}^\infty (\phi_{vs} - \phi_{ls})dy
\end{equation}
If $\delta_0$ is the smallest distance between particles of the substrate and fluids when
they are in contact, Eq.~(\ref{eq:energy_diff}) specifies the total energy required to
completely remove the liquid from the substrate.  Even if $\delta_0$ takes a larger value,
Eq.~\eqref{eq:energy_diff} consistently describes the energy difference between a wet
system, and a dry system which consists of a fluid layer of thickness $\delta_0$ wetting
the solid substrate.

Performing the integration in Eq.~\eqref{eq:energy_diff}, we obtain:
\begin{equation}\label{eq:spread_micro_simp}
	S_{eq} = (\mathcal{K}_{vs}-\mathcal{K}_{ls})h^\ast\left[\frac{1}{m-1}\left( \frac{h^\ast}{\delta_0} \right)^{m-1} - \frac{1}{n-1}\left( \frac{h^\ast}{\delta_0} \right)^{n-1}\right]
\end{equation}
In order to be in a partial wetting regime, i.e.~where there is a non-zero contact angle,
we require $S_{eq}<0$. Based on Eq.~\eqref{eq:spread_micro_simp}, we find:
\begin{enumerate}
 \item $\delta_0 > \left(\frac{n-1}{m-1}\right)^{\frac{1}{m-n}}h^\ast$:
	In this case,
	the attractive term in Eq.~\eqref{eq:spread_micro_simp} dominates.
	This leads to a negative spreading coefficient only if $\mathcal{K}_{vs}>\mathcal{K}_{ls}$,
	so that the vapor phase must experience a greater attraction than the liquid phase
	in order to be in a partial wetting regime.
\item $\delta_0 < \left(\frac{n-1}{m-1}\right)^{\frac{1}{m-n}}h^\ast$: In this case,
	the repulsive term in Eq.~\eqref{eq:spread_micro_simp} dominates.
	Thus partial wetting is possible when $\mathcal{K}_{vs}<\mathcal{K}_{ls}$.
\end{enumerate}
The first case states that when $\delta_0$ is large, the vapor phase must interact more
strongly with the substrate than the liquid phase; the reverse is true when $\delta_0$ is
small.  For example, the second case must hold for partial wetting of a droplet surrounded
by a vacuum. 

For comparison, we briefly describe a similar method of contact angle implementation used
in the long-wave theory~\cite{Schw98,dk_pof07}.  In its most basic form, the disjoining
pressure modifies the pressure jump across the interface of a thin, flat film, and can be
derived as a macroscopic consequence of the van der Waals interaction~\cite{carey99}.  For
a flat film of thickness $h$, the difference between the pressure in the exterior vapor
phase, $p_v$, and the liquid phase, $p_l$, is given by:
\begin{equation}\label{eq:jump_dj}
	p_v - p_l = \Pi(h) = K_{dj}\left[\left(\frac{h^\ast}{h}\right)^m -\left(\frac{h^\ast}{h}\right)^n \right].
\end{equation}
The definition of the parameter $h^\ast$ is identical to that of
Eq.~\eqref{eq:spread_micro_simp}.  In the long-wave literature, this quantity is
identified with an equilibrium film thickness, such that in the case of a drop in the
partial wetting regime, there is an additional   film of thickness $h^\ast$ completely
wetting the substrate.

Following the example of disjoining pressure above, we consider situations such that there
is a layer of thickness $h^\ast$ completely covering the surface even when the fluid is
partially wetting.  In particular, this means that we assume that $\delta_0 = h^\ast$, so
that in Eq.~\eqref{eq:energy_diff}, the liquid is removed only to a thickness $h^\ast$.
This value of $\delta_0$ has the convenient property that it allows for
Eq.~\eqref{eq:spread_micro_simp} to be simplified, while the presence of a wetting layer
over the whole substrate removes the contact line singularity.  Plugging $\delta_0=h^\ast$
into Eq.~\eqref{eq:spread_micro_simp}, and substituting in the expression for $S_{eq}$
from Eq.~\eqref{eq:spreading_II}, we obtain the following expression for the difference
between $\mathcal{K}_{vs}$ and $\mathcal{K}_{ls}$ as a function of $\theta_{eq}$:
\begin{equation}\label{eq:form_for_K}
	\mathcal{K}_{vs} - \mathcal{K}_{ls}=\frac{\gamma(1-\cos\theta_{eq})}{h^\ast}\left(\frac{(m-1)(n-1)}{m-n}\right)
\end{equation}
Therefore, only the difference $\mathcal{K}_{vs}-\mathcal{K}_{ls}$ is needed to specify
$\theta_{eq}$.  Similar formulations appear in the long-wave literature~\cite{Schw98,dk_pof07}.  
Note that since Eq.~\eqref{eq:energy_diff} assumes that the fluid occupies a
half infinite domain in the wet state, Eq.~\eqref{eq:form_for_K} is satisfied exactly only
in the limit where $h^\ast$ is vanishingly small relative to the drop thickness.

To summarize, in this section we have formulated a method that allows for the inclusion of
fluid/solid interaction forces without the limitations inherent in long-wave theory, such
as negligible inertia and small interfacial slope.   Next we proceed to discuss numerical
implementation.

\section{Numerical Methods}\label{sec:methods}
Equations~\eqref{eq:interaction} and ~\eqref{eq:force} present a difficulty in that they
diverge as $y\rightarrow 0$. This can be dealt with by introducing a cutoff length,
i.e.~forcing $\textbf{F}_{is}=0$ when $y$ is less than the cutoff.  This method has the
drawback of making the force discontinuous, and may lead to poor numerical convergence.
For this reason, we introduce a shifted potential:
\begin{equation}\label{eq:pot_shifted}
	\hat{\Phi}_{is} = \mathcal{K}_{is}\left[ \left(\frac{h^\ast}{y+h_c}\right)^m - \left(\frac{h^\ast}{y+h_c}\right)^n \right]
\end{equation}
where $h_c$ is some parameter less than $h^\ast$.
Equation~\eqref{eq:pot_shifted} removes the singular portion of the potential
near $y\rightarrow 0$ for a small $h_c$, and is equal to zero at
$y=h^\ast-h_c$, but otherwise keeps the same functional form.  We refer to
films occupying $0\leq y\leq h^\ast-h_c$ as equilibrium films, and $h^\ast-h_c$
as the equilibrium film thickness.  The force corresponding to
Eq.~\eqref{eq:pot_shifted} is given by
\begin{equation}\label{eq:force_shifted}
	\hat{\textbf{F}}_{is} = \frac{\mathcal{K}_{is}}{h^\ast}\left[m\left( \frac{h^\ast}{y+h_c} \right)^{m+1} - n\left(\frac{h^\ast}{y+h_c} \right)^{n+1} \right]\hat{y}
\end{equation}
We can express the total body force due to the van der Waals interaction more
compactly by introducing a characteristic function, $\chi$, which takes the
value $1$ inside the liquid phase and $0$ elsewhere: 
\begin{equation}\label{eq:total_body_force}
	\hat{\textbf{F}}_{B}(y) = \chi\hat{\textbf{F}}_{ls} + (1-\chi)\hat{\textbf{F}}_{vs}
\end{equation}
where the interaction strength explicitly depends on the phase through $\chi$, so that
$\mathcal{K}(1) = \mathcal{K}_{ls}$ and $\mathcal{K}(0) = \mathcal{K}_{vs}$.

We solve the full two-phase Navier-Stokes equations subject to the force given by
Eq.~\eqref{eq:force_shifted}:
\begin{multline}\label{eq:NS}
	\rho(\chi) (\partial_{t} \textbf{u} + \textbf{u}\cdot\nabla\textbf{u}) = -\nabla p + \nabla\cdot\left[ \mu(\chi)(\nabla\textbf{u} + \nabla \textbf{u}^T) \right] + \gamma \kappa \delta_s \textbf{n} 
		\\+ \hat{\textbf{F}}_B(y)
\end{multline}
\begin{equation}\label{eq:continuity}
\nabla \cdot \textbf{u} = 0
\end{equation}
The phase dependent density and viscosity are given by $\rho(\chi) = \chi\rho_l
+ (1-\chi)\rho_v$ and $\mu(\chi) = \chi\mu_l + (1-\chi)\mu_v$.  The velocity
field is $\textbf{u}=(u,v,w)$, and $p$ is the pressure.  The force due to the
surface tension is included as a singular body force in the third term on the
right hand side of Eq.~\eqref{eq:NS} (see~\cite{Brackbill92}), where $\delta_s$
is a delta function centered on the interface, $\kappa$ is the curvature, and
$\textbf{n}$ is the inward pointing unit normal of the interface.  Letting $L$
be the length scale, and $\tau$ the time scale, we define the following
dimensionless variables:
\begin{eqnarray*}
 \tilde{x} = \frac{x}{L} 	&&  	\tilde{y} = \frac{y+h_c}{L}\\
 \tilde{z}=\frac{z}{L} 		&&	\tilde{t}=\frac{t}\tau	\\	
 \tilde{h}^\ast = \frac{h^\ast}{L} && 	\tilde{h}_c = \frac{h_c}{L}\\
 \tilde{\textbf{u}} = \frac{\textbf{u}\tau}{L} && \tilde{p} = \frac{L}{\gamma}p\\
 \tilde{\kappa} = L\kappa	&&	\tilde{\delta_s} = L\delta_s\\
\end{eqnarray*}
With these scales, and dropping the tildes, the dimensionless Navier-Stokes equations are
as follows:
\begin{multline}\label{eq:NS_nondim}
	We(\chi) (\partial_{t} \textbf{u} + \textbf{u}\cdot\nabla\textbf{u}) = -\nabla p + \nabla\cdot\left[ Ca(\chi)(\nabla\textbf{u} + \nabla \textbf{u}^T) \right] + \kappa \delta_s \textbf{n} \\
		+   \textbf{F}_B(y)
\end{multline}
\begin{equation}\label{eq:divergencefree}
	\nabla\cdot \textbf{u} = 0 
\end{equation}
where
\begin{equation}\label{eq:nondim_F}
	\textbf{F}_B(y)=K(\chi)\mathcal{F}(y)\hat{y} = K(\chi)\left[m\left(\frac{h^\ast}{y}\right)^{m+1} - n\left(\frac{h^\ast}{y}\right)^{n+1}\right]\hat{y}
\end{equation}
Here, the dimensionless $y$-coordinate is translated, so that
$\mathcal{F}(h^\ast)=0$, and the equilibrium film occupies $h_c\leq y\leq
h^\ast$.  The (phase-dependent) dimensionless numbers are the Weber number,
$We$, the capillary number, $Ca$, and the force scale for the van der Waals
interaction, $K$, given by the following:
\begin{align*}
		We(\chi) &=& \frac{\rho(\chi) L^3}{\tau^2 \gamma}  &=& \chi We_l + (1-\chi)We_v  &=& \frac{L^3}{\tau^2\gamma}\left(\chi \rho_l + (1-\chi)\rho_v \right)  \\
		Ca(\chi) &=& \frac{\mu(\chi) L}{\gamma \tau} &=& \chi Ca_l + (1- \chi)Ca_v\  &=& \frac{L}{\gamma \tau} \left(\chi\mu_l + (1- \chi)\mu_v\right)  \\
		K(\chi) &=& \frac{\mathcal{K}(\chi)L}{h^\ast \gamma} &=& \chi K_{ls} + (1-\chi) K_{vs} &=& \frac{L}{h^\ast \gamma} \left(\chi \mathcal{K}_{ls} + (1-\chi) \mathcal{K}_{vs}   \right)
\end{align*}
Upon nondimensionalization, Eq.~\eqref{eq:form_for_K} yields the following expression for
$\theta_{eq}$: 
\begin{equation}\label{eq:nondim_K}
	\Delta K(\theta_{eq}) := K_{vs} - K_{ls} = \frac{(1-\cos\theta_{eq})}{ {h^\ast}^2 } \left( \frac{(m-1)(n-1) }{(m-n) }\right)
\end{equation}

\begin{figure}
 \centering
 \includegraphics{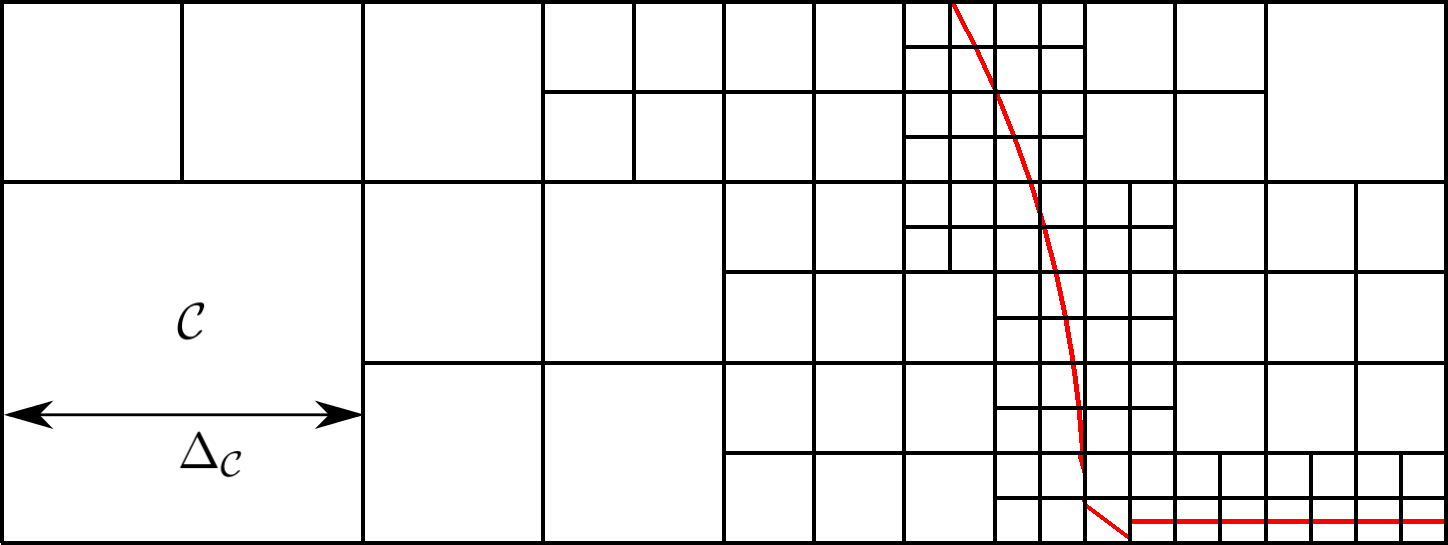}
 \caption{Illustration of the quad-tree used to discretize the continuity
   domain. The mesh resolution can be chosen adaptively, so that each cell
   $\mathcal{C}$ has a corresponding size $\Delta_{\mathcal{C}}$. The red curve
   shows the piecewise linear reconstruction of the liquid/vapor interface
   obtained using the volume of fluid method.}
 \label{fig:tree_diagram}
\end{figure}

The system of Eqs.~\eqref{eq:NS_nondim}-\eqref{eq:divergencefree}, is solved
using the open-source package Gerris~\cite{Popinetgerris}, described in detail
in~\cite{Popinet2009}.  In Gerris, the Navier-Stokes equations are solved using
a finite-volume based projection method, with the interface between the two
fluids tracked using the volume of fluid method.  The fluid domain is
discretized into a quad-tree of square cells (see Fig.~\ref{fig:tree_diagram});
note that we consider the implementation in two spatial dimensions from now
onward.  The finite-volume method treats each cell $\mathcal{C}$ as a control
volume, so that the variables associated with $\mathcal{C}$ represent the
volume averaged value of the variable over $\mathcal{C}$. Each cell has a size
$\Delta_{\mathcal{C}}$, and center coordinates $(x_c, y_c)$.  The volume of
fluid method tracks the interface implicitly by defining a volume fraction
function, $T(\mathcal{C})$, which gives the fraction of the volume of
$\mathcal{C}$ occupied by the liquid phase. $T(\mathcal{C})$ is advected by the
fluid flow according to the following equation:
\begin{equation}\label{eq:vodadvection}
 \partial_t T + \nabla\cdot (\textbf{u} T) = 0
\end{equation}
The value of $We$ and $Ca$ in a cell $\mathcal{C}$ are then represented using an average:
\begin{align*}
	We(\mathcal{C}) &= We_lT(\mathcal{C})  + We_v (1 - T(\mathcal{C}))\\
	Ca (\mathcal{C}) &= Ca_l T(\mathcal{C}) + Ca_v (1 - T(\mathcal{C}))\\		
\end{align*}
Here, as before, subscript $l$ corresponds to liquid, and $v$ to the vapor phase.

The volume of fluid method reconstructs a sharp interface.  For each cell, the
gradient, $\nabla T$, is computed using the Mixed-Young's
method~\cite{Aulisa2007}, so that the unit normal vector is given by
$\textbf{M}=\nabla T/|\nabla T|$.  This permits a linear reconstruction of the
interface in each cell, according to the equation $\textbf{M}\cdot \textbf{x} =
\alpha$, where $\alpha$ is determined by $T$ (see~\cite{scardo2000} for
details).

The exact average of the force over the computational cell, $\mathcal{C}$, is given by:
\begin{multline}\label{eq:discrete_exact}
	\textbf{F}_{B}(\mathcal{C}) = \frac{1}{\Delta_C^2} \int \int_{\mathcal{C}} K(\chi(x,y))\mathcal{F}(y) d\mathcal{C}\,
	\hat{y} \\= \frac{1}{\Delta_C^2} \int \int_{\mathcal{C}} \left[K_{ls}\chi(x,y) + K_{vs}(1-\chi(x,y))\right]\mathcal{F}(y) d\mathcal{C}\,\hat{y}
\end{multline}
where $\mathcal{F}(y)$ is defined by Eq.~\eqref{eq:nondim_F}. We detail two possibilities
for the discretization of the van der Waals force term in Eq.~\eqref{eq:NS_nondim}.  For
both of these methods, the force is included explicitly in the predictor step of the
projection method.

\textbf{Method I:} 

Our first method proceeds by a simple second order discretization of
Eq.~\eqref{eq:discrete_exact}:
\begin{equation*}
	\begin{split}
	\textbf{F}_I (\mathcal{C}) 
	&= \frac{\mathcal{F}(y_c)}{\Delta_C^2}\int\int_\mathcal{C} \left[K_{ls}\chi(x,y) + K_{vs}(1-\chi(x,y))\right] d\mathcal{C}\,\hat{y}\\
	&= \left[\frac{\mathcal{F}(y_c)}{\Delta_C^2}K_{ls} \int\int_\mathcal{C} \chi(x,y)d\mathcal{C} 
			+ \frac{\mathcal{F}(y_c)}{\Delta_C^2}K_{vs} \int\int_\mathcal{C} (1-\chi(x,y))d\mathcal{C}\right]\,\hat{y}\\
	\end{split}
\end{equation*}
We identify the average of $\chi$ over $\mathcal{C}$ with $T(\mathcal{C})$, yielding the
following expression:
\begin{equation}\label{eq:disc_meth_I}
	\textbf{F}_I (\mathcal{C}) = 
	  \left[T(\mathcal{C})K_{ls} + (1-T(\mathcal{C}))K_{vs}\right]
	    \mathcal{F}(y_c)\,\hat{y}
\end{equation}
However, we note that the accuracy of this simple method can deteriorate at low mesh
resolutions because $\mathcal{F}$ has a large gradient as $y\rightarrow 0$.  Consider a
cell such that $T(\mathcal{C})=1$. To a first approximation, the error in
Eq.~\eqref{eq:disc_meth_I} is given by:
\begin{multline*}
  \mathcal{E}=|\textbf{F}_{B}(\mathcal{C}) - \textbf{F}_I(\mathcal{C})| = \left| K_{ls}\frac{\Delta_C^2}{24}\mathcal{F}''(y_c) + o(\Delta_C^2)\right|\\
  \approx \frac{\Delta_C^2}{24}\frac{K_{ls}}{{h^\ast}^2}\left| m(m+1)(m+2)\left(\frac{h^\ast}{y}\right)^{m+3} \right.\\
	\left.- n(n+1)(n+2)\left(\frac{h^\ast}{y}\right)^{n+3}\right|
\end{multline*}
To show that this can lead to large errors, consider the error when $\Delta_\mathcal{C} =
h^\ast-h_c$, in a cell with center at $y_c = (h^\ast+h_c)/2$, i.e.~for a cell along the
bottom boundary when $h^\ast$ is just barely resolved.  The error in this cell is:
\begin{multline*}
	\mathcal{E}_1 = \frac{(1-h_c/h^\ast)^2}{24}K_{ls}\left| m(m+1)(m+2)\left(\frac{2}{1+h_c/h^\ast}\right)^{m+3}  \right. \\
		\left. - n(n+1)(n+2)\left(\frac{2}{1+h_c/h^\ast}\right)^{n+3} \right|
\end{multline*}
The lower bound of this error can be shown to be:
$$
	\mathcal{E}_1 > \frac{(1-h_c/h^\ast)^2}{24}K_{ls}\left(m(m+1)(m+2) - n(n+1)(n+2)\right)
$$
Since $m \geq n+1$, we can bound this further by:
\begin{equation}\label{eq:meth_I_error}
	\mathcal{E}_1 > \frac{(1-h_c/h^\ast)^2}{8}K_{ls}m(m+1)
\end{equation}
Note that $\mathcal{E}_1$ is not the discretization error, but the error found
for a fixed grid size.

Recalling that Eq.~\eqref{eq:nondim_K} implies that $K_{ls}$ is proportional to
$1/{(h^\ast)}^{2}$, so if $h_c$ is significantly smaller than $h^\ast$ the
errors are quite large unless $\Delta_C$ is much smaller than $h^\ast$.
Furthermore, $h^\ast$ is generally small relative to the length scale of the
droplet considered in Eq.~\eqref{eq:spread_micro_simp}, so that the mesh size
required to accurately resolve the right hand side of
Eq.~\eqref{eq:discrete_exact} will increase the computational cost.

\begin{figure}
 \centering
 \begin{subfigure}{0.4\textwidth}
	 \centering
 	\includegraphics[width=2in]{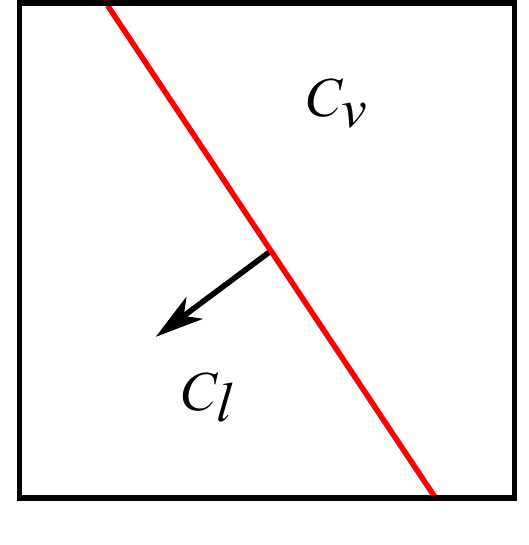} 
	\caption{}
	\label{fig:cut_cell}
 \end{subfigure}
 \begin{subfigure}{0.4\textwidth}
	 \centering
 	\includegraphics[width=2in]{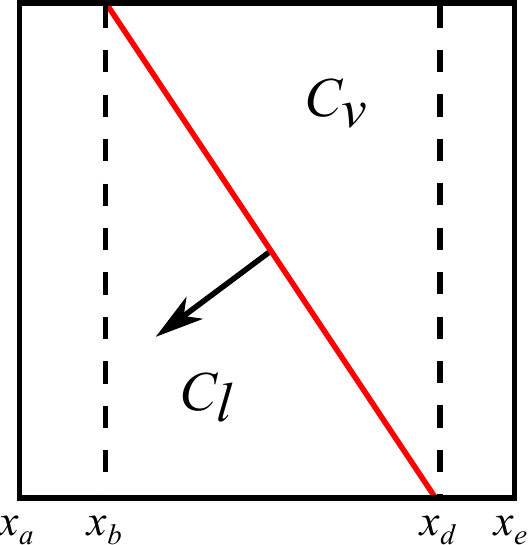}
	\caption{}
	\label{fig:cut_cell_int}
 \end{subfigure}
 \caption{Illustration of a cell with $0<T(\mathcal{C}) < 1$: (a) cut cell
   divided into regions $C_l$ and $C_v$ entirely occupied by the liquid and vapor
   phases, respectively, and (b) illustration of the regions of integration of a
   cut cell. The linear reconstruction of the interface from the volume of fluid
   method is shown by the red line.}
 \label{fig:cell_cartoon}
\end{figure}

\textbf{Method II:} 

Since Eq.~\eqref{eq:nondim_F} gives an exact formula for the force per unit
volume in a single phase, much of the simplification used in deriving
Eq.~\eqref{eq:disc_meth_I} is unnecessary. For the second method, we take
advantage of this fact to more accurately approximate
Eq.~\eqref{eq:discrete_exact}.  First, note that it is trivial to integrate
Eq.~\eqref{eq:discrete_exact} when $T(\mathcal{C})=1$ or $T(\mathcal{C})=0$.
Moreover, the volume of fluid method gives additional information beyond just
the fraction of the cell occupied by the liquid phase; using the reconstructed
interface, we have an approximation of the portion of the cell which is
occupied by the liquid phase as well.  Therefore, we introduce the following
alternative method.  Consider a cell $\mathcal{C}$ with center $(x_c, y_c)$,
where $0<T(\mathcal{C})<1$, so that the volume of fluid method yields a linear
reconstruction of the interface in the cell.  This interface divides the cell
into a region occupied by the liquid phase, $C_l$, and a region occupied by the
vapor phase, $C_v$ (see Fig.~\ref{fig:cut_cell}).  We can write a general
algorithm to integrate over these regions, yielding the following expression
for the average force on cell $\mathcal{C}$:
\begin{equation}\label{eq:disc_meth_II}
	\textbf{F}_{II}(\mathcal{C}) = \left[
			\frac{\mathcal{K}_{ls}}{\Delta_C^2} \int\int_{C_l} \mathcal{F}(y)dxdy + 
			\frac{\mathcal{K}_{vs}}{\Delta_C^2} \int\int_{C_v} \mathcal{F}(y)dxdy
			\right]\hat{y}
\end{equation}
We briefly outline the general process to perform the integrations over $C_l$
and $C_v$.  Let the volume of fluid reconstructed interface in the cell be
given by $\textbf{M}\cdot \textbf{x} = \alpha$, where $\textbf{M} = (M_x, M_y)$
is a normal pointing into the liquid phase.  We only present the problem of
integrating over $C_l$, since $C_v$ involves an analogous procedure.  If
$|M_x|<\xi$, for a small tolerance $\xi$, the interface is approximately
horizontal; thus the integration of $\mathcal{F}(y)$ over $C_l$ becomes:
$$
	\frac{1}{\Delta_C^2}\int\int_{C_l} \mathcal{F}(y) dx dy 
	  = \frac{1}{\Delta} \int_{y_c-\Delta/2}^{y_c-\Delta/2+T\Delta}\mathcal{F}(y)dy
$$
On the other hand, if $|M_y|<\xi$, the interface is approximately vertical, and the
integration over $C_l$ becomes:
$$
	\frac{1}{\Delta_C^2}\int\int_{C_l} \mathcal{F}(y) dx dy = 
	  \frac{T}{\Delta} \int_{y_c-\Delta/2}^{y_c+\Delta/2}\mathcal{F}(y)dy
$$
We can now proceed to provide general formulas when $|M_y| > \xi$ and $|M_x| > \xi$.
Since $\mathcal{F}(y)$ is independent of $x$, the sign of the interfacial slope is
irrelevant, so, without loss of generality, suppose that $M_x/M_y > 0$.  Define the
following values:
\begin{align*}
	x_a &= x_c-\Delta_C/2\\
	x_b &= \max \left( \frac{-M_y(y_c+\Delta_C/2)+\alpha}{M_x} , x_c-\Delta_C/2\right)\\
	x_d &= \min\left(  \frac{-M_y(y_c-\Delta_C/2)+\alpha}{M_x},  x_c+\Delta_C/2\right)\\
	x_e &= x_c+\Delta_C/2
\end{align*}
If $M_y<0$, the integration over $C_l$ can be expressed as follows:
\begin{multline}\label{eq:methII_my_lt_0}
		\frac{1}{\Delta_C^2}\int\int_{C_l} \mathcal{F}(y) dx dy =\\ 
			\frac{1}{\Delta_C^2}\left[
			 (x_b-x_a)\int_{y_c-\Delta/2}^{y_c+\Delta/2}\mathcal{F}dy
			+ \int_{x_b}^{x_d} \int_{y_c-\Delta/2}^{-M_x/M_y x + \alpha/M_y} \mathcal{F}dydx \right]
\end{multline}
On the other hand, if $M_y >0$, the integration over $C_l$ is given by:
\begin{multline}\label{eq:methII_my_gt_0}
		\frac{1}{\Delta_C^2}\int\int_{C_l} \mathcal{F}(y) dx dy = \\
		\frac{1}{\Delta_C^2}\left[
			\int_{x_b}^{x_d} \int_{-M_x/M_y x + \alpha/M_y}^{y_c+\Delta/2} \mathcal{F}dydx
			+ (x_e - x_d) \int_{y_c-\Delta/2}^{y_c+\Delta/2}\mathcal{F}dy
			\right]
\end{multline}
Thus, in two dimensions, the general process of integration reduces to integration over
two regions.  Note again that $\mathcal{F}(y)$ is known exactly, so that the integrals in
Eqs.~\eqref{eq:methII_my_lt_0}-\eqref{eq:methII_my_gt_0} can be computed exactly, in
contrast to the large error for Method I in Eq.~\eqref{eq:meth_I_error}.  Therefore, we
conclude that regarding approximation of Eq.~\eqref{eq:discrete_exact}, Method II is
superior.  For example, for a single phase fluid, or when the interface is flat,
Eq.~\eqref{eq:disc_meth_II} is exact.  We will consider both Method I and Method II in
what follows and discuss their performance.

Our computational domain is rectangular, with $x\in(0,x_{max})$ and $y \in (h_c,
y_{max})$.  Throughout the remainder of the paper, we will impose a homogeneous Neumann
boundary condition on all boundaries for the pressure, $p$.  For the velocity field, we
will apply a homogeneous Neumann boundary condition at all boundaries except $y=h_c$,
where we will apply one of the following two boundary conditions; the no-slip,
no-penetration condition by setting ${\bf u}$ to $0$ on the substrate:
$$
	(u,v)|_{y=h_c} = (0,0)
$$
or a free-slip condition by:
\begin{eqnarray*}
	\partial_y u|_{y=h_c} = 0 & v|_{y=h_c} =0
\end{eqnarray*}
The boundary condition for the van der Waals force is straightforward. 
At $y=h_c$, we set
$$
	\textbf{F}_{I,II} |_{y=h_c} = K_{ls}\mathcal{F}(h_c)
$$
At $y=y_{max}$, we set
$$
	\textbf{F}_{I,II} | _{y=y_{max}} = K_{vs}\mathcal{F}(y_{max})
$$
The boundary condition for the volume fraction, $T$, is again homogeneous Neumann on all
boundaries, except on the bottom boundary, where we apply
$$
	T |_{y=h_c} = 1
$$
which is equivalent to taking the bottom boundary to always be wetted with the liquid
phase.

\section{Results}\label{sec:results}
In this section, we consider simulations of the full Navier-Stokes equations in
which contact angles have been imposed using the van der Waals force, which is
included using Methods I and II described in Sec.~\ref{sec:methods}.  Our
simulation setup consists of a drop on an equilibrium film, initially at rest,
which then relaxes to equilibrium under the influence of the van der Waals
force.  We compare Methods I and II for a drop which is initially close to its
equilibrium contact angle, as predicted by Eq.~\eqref{eq:nondim_K}, as well as
for a spreading drop which is initially far from its equilibrium.  Furthermore,
we consider the effect of equilibrium film thickness for small, intermediate,
and large contact angles, using Method II.

In simulations, droplets are surrounded by an equilibrium film of thickness
$\bar{h^\ast}:=h^\ast-h_c$ defined by $y<h^\ast$, where $h_c$ is the amount the force is
translated by in Eq.~\eqref{eq:force_shifted}. The initial profile is given by the
following:
\begin{equation}\label{eq:initial_condition}
 	\left\{ (x,y) : x^2+(y+R\cos\theta_i-h^\ast)^2 < R^2 \text{ or } y<h^\ast\right\}
\end{equation}
Here $R$ is chosen so that the total area of the circular cap is equal to
$A_0=\pi0.75^2/2$.  The quantity $\theta_i$ is the initial contact angle of the
droplet.  In all simulations, we fix the ratio of these length scales so that
$h^\ast=2h_c$. Various values of the exponents $m$ and $n$ in
Eq.~\eqref{eq:nondim_F} can be found in the literature, in particular $(m,n) =
(9,3),~(4,3),~(3,2)$ have been used in the context of the disjoining pressure
in thin films~\cite{cm_rmp09,gonzalez2013}, the former corresponding to
the 12-6 Lennard Jones potential.  In this paper we will restrict our attention
to $(m,n)=(3,2)$, which was shown to lead to favorable agreement with
experiments involving liquid metal films~\cite{gonzalez2013}.  Different
exponents may affect the structure of the contact line region and the pressure
distribution, but do not impact significantly the results presented here.

There are four contact angles that we consider in this section.  The initial
contact angle, $\theta_i$, as in Eq.~\eqref{eq:initial_condition}, specifies
the angle formed by the tangent of the initial circular profile with the
equilibrium film at the point where they meet.  The (time dependent) apparent
contact angle is denoted $\theta$, and the numerical equilibrium angle
$\theta_{num}$ is the value of $\theta$ when the system under consideration is
in equilibrium; $\theta$ is measured by a circle fit procedure described below.
Finally, we refer to the imposed contact angle, $\theta_{eq}$, which we use to
specify $K_{vs}-K_{ls}$ via Eq.~\eqref{eq:nondim_K}; the only relevant quantity
is the difference between $K_{vs}$ and $K_{ls}$, but for definiteness, we set
$K_{vs}=1.1\Delta K(\theta_{eq})$ and $K_{ls} = 0.1\Delta K(\theta_{eq})$.  The
value of $\theta_{eq}$ differs from $\theta_{num}$ in simulations because
$h^\ast$ is small but non negligible relative to the thickness of the drop.

We measure $\theta$ according to the following procedure.  A drop has a
circular cap shape that transitions smoothly to the equilibrium film.  We
define the contact angle in this context in the following way: let $\pm
x_{inf}$ represent the points of inflection of the drop profile.  We perform a
least squares fit of a circle to the profile over the interval $(-x_{inf},
x_{inf})$.  The point at which the fitted circle intersects the equilibrium
film is called the contact point (in three dimensions, this is the contact
line).  $\theta$ is measured as the angle the fitted circle makes with the
equilibrium film at the contact point.

We first investigate a static drop, with $h^\ast=0.03$ and $\theta_i=\pi/2$,
which is then allowed to relax to its equilibrium shape. The dimensionless
numbers are set to be $We_l=We_v=Ca_l=Ca_v=0.05$.  We set the material
parameters to be the same in both fluid phases in order to restrict our
attention to a smaller parameter space, so that we can focus on the properties
of Methods I and II.  Note that since the expression determining $\theta_{eq}$,
Eq.~\eqref{eq:nondim_K}, does not depend on the fluid parameters, this choice
does not affect the equilibrium shape.  The computational domain is $(x,y) \in
[0,2] \times [h_c,1+h_c]$; the domain is resolved at a constant resolution.  A
no-slip boundary condition is imposed on the solid boundary, $y=h_c$.
\begin{figure}[t]
 \includegraphics{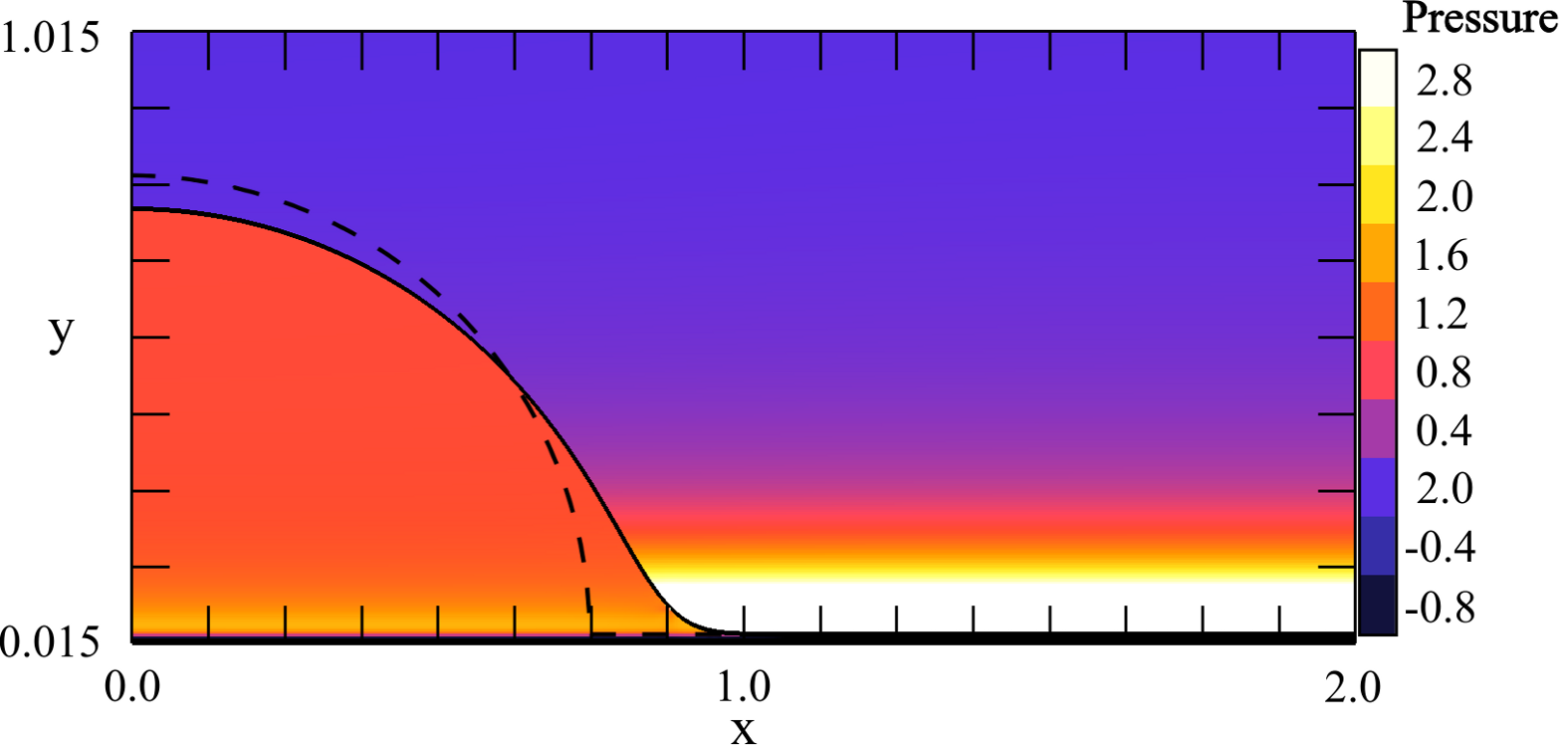}
 \caption{Pressure distribution for a drop with $\theta_{eq}=\pi/2$,
	 $h^\ast=0.03$, at equilibrium. The pressure is normalized
	  by the (non-dimensional) capillary pressure of the equilibrium drop,
	  which is given by the curvature of the interface far from the contact
	  point, where it is approximately circular. 
	  The solid black curve shows the volume of fluid reconstructed
	  interface, and the dashed curve is the initial profile; $\Delta=1/256$.}
 \label{fig:eq_drop_profile}
\end{figure}
Figure~\ref{fig:eq_drop_profile} shows such a drop at equilibrium (with the
contact angle imposed using Method II).  The equilibrium film has negative
pressure with very high absolute value, and its thickness at equilibrium
differs from $\bar{h^\ast}$ by about $3\%$.  With this $h^\ast$, there is a
noticeable difference between the true equilibrium contact angle found by
simulations, $\theta_{num}$, and the imposed angle $\theta_{eq}$.  The effect
of $h^\ast$ on the equilibrium drop shape will be considered below.

\begin{figure}[t]
 \centering
 \includegraphics{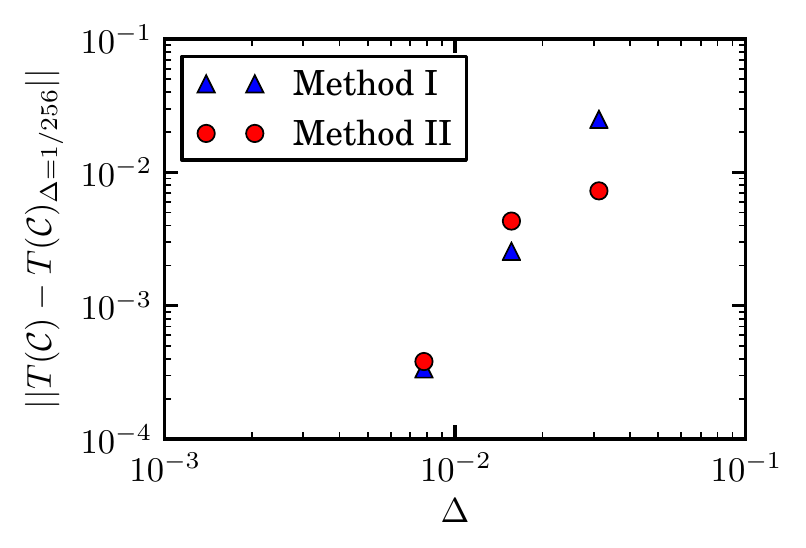}	
 \caption{The convergence of the $L^1$ norm of the difference between
   $T(\mathcal{C})$ at $\Delta$ and $\Delta=1/256$, for a drop at equilibrium
   with $\theta_{eq}=\pi/2$, $h^\ast=0.03$, with a uniform mesh size $\Delta$.}
 \label{fig:l1_error}
\end{figure}

Figure~\ref{fig:l1_error} shows the convergence for Methods I and II for a uniform mesh.
Simulations are run until time $t=1.75$, with a constant time step of $\Delta t\approx
2.8\times10^{-5}$; the stability constraint due to the explicit discretization of the
surface tension dominates in this velocity regime, and this time step ensures that this
constraint is satisfied for all resolutions we consider (see~\cite{Popinetgerris}).  For
two sets of volume fractions $T_1$ and $T_2$ such that $T_2$ is computed on a quad-tree
$Q$ with mesh-size $\Delta_Q$, the $L^1$ norm is computed according to the following
formula: 
\begin{equation}\label{eq:l1norm} 
  ||T_1(\mathcal{C}) -T_2(\mathcal{C})|| =
    \sum_{\mathcal{C} \in Q} |T_1(\mathcal{C}) - T_2(\mathcal{C})| \Delta_Q^2 
  \end{equation}
In this case $\Delta_Q=1/256$, and $T_1(\mathcal{C})$ is equal to $T_1$ on the largest
cell containing $\mathcal{C}$.  For reference, $\Delta=1/32$ is about $h^\ast$, so that
the low resolution case is not even resolving the equilibrium film.  Both methods perform
comparably well at $\Delta=1/64, 1/128$, with Method I converging slightly faster.
However, at low resolutions, $\Delta=1/32$, Method II performs significantly better.  

\begin{figure}
 \centering
 \includegraphics{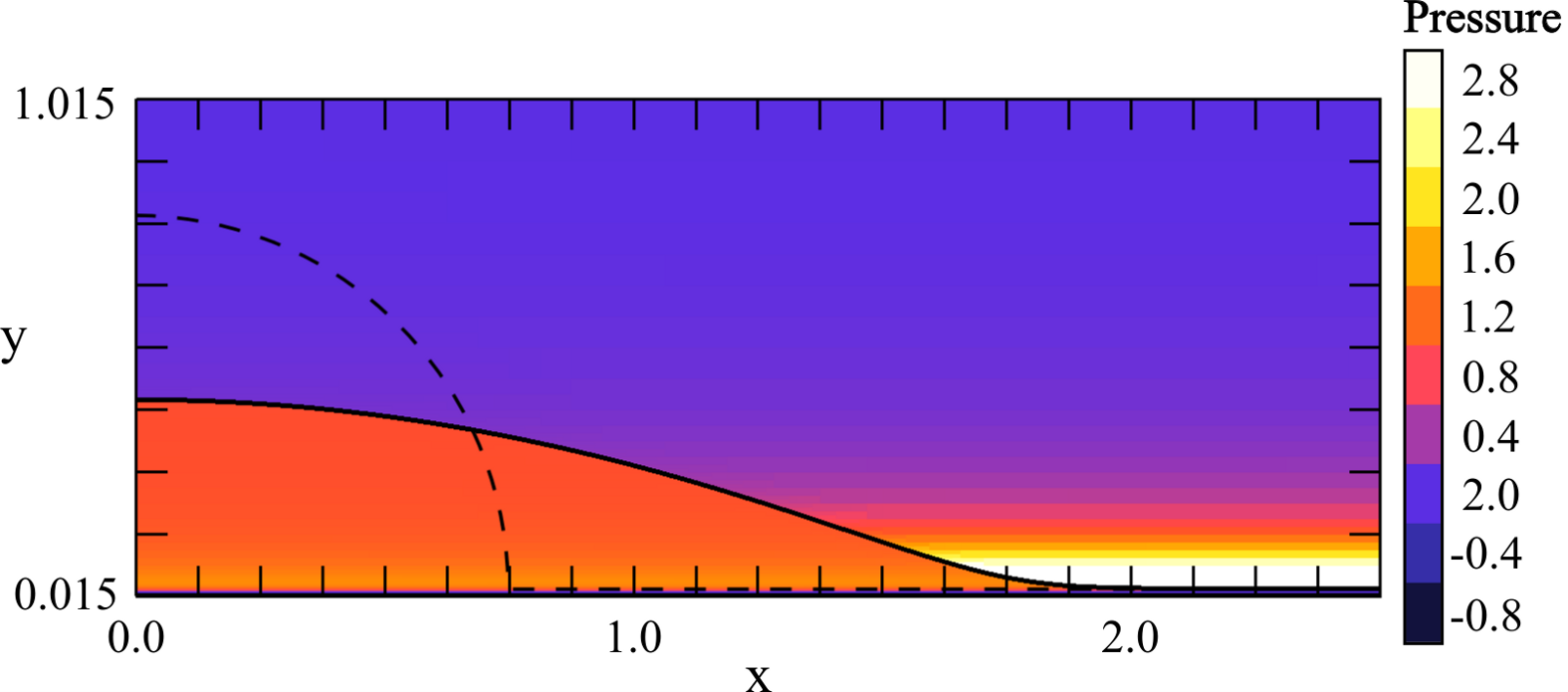}
 \caption{Pressure distribution for a drop at equilibrium with
   $\theta_{eq}=\pi/6$.  Initially, $\theta_i=\pi/2$, and the drop spreads to
   its equilibrium configuration, defined by $\theta_{eq} = \pi/ 6$.  The
   initial profile is shown by the dashed curve.  As in
   Fig.~\ref{fig:eq_drop_profile}, the pressure is normalized by the capillary
   pressure of the equilibrium drop; $\Delta=1/256$.}
 \label{fig:spread_profile}
\end{figure}

\begin{figure}
 \centering
 \begin{subfigure}\textwidth
  \includegraphics[width=\textwidth]{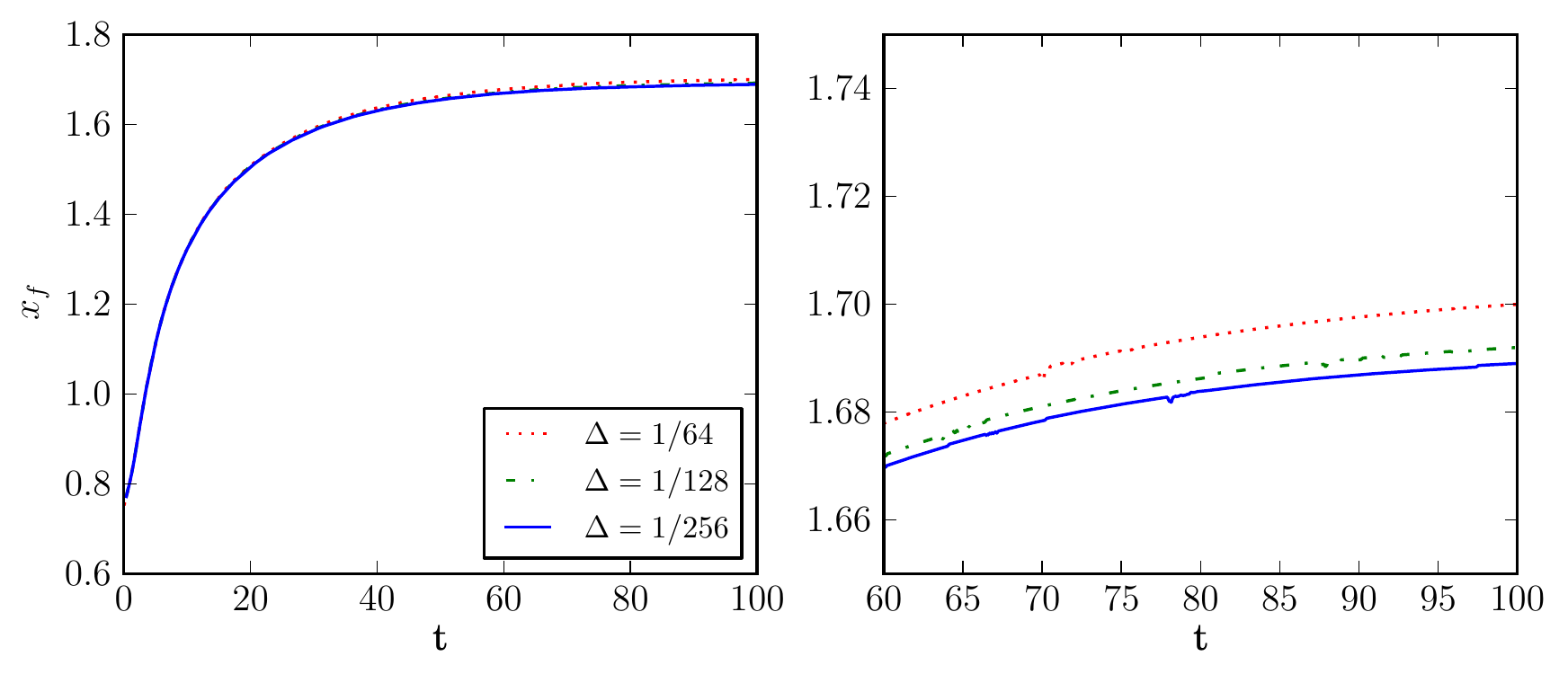}
  \caption{Method I}
 \end{subfigure}
 \begin{subfigure}\textwidth
  \includegraphics[width=\textwidth]{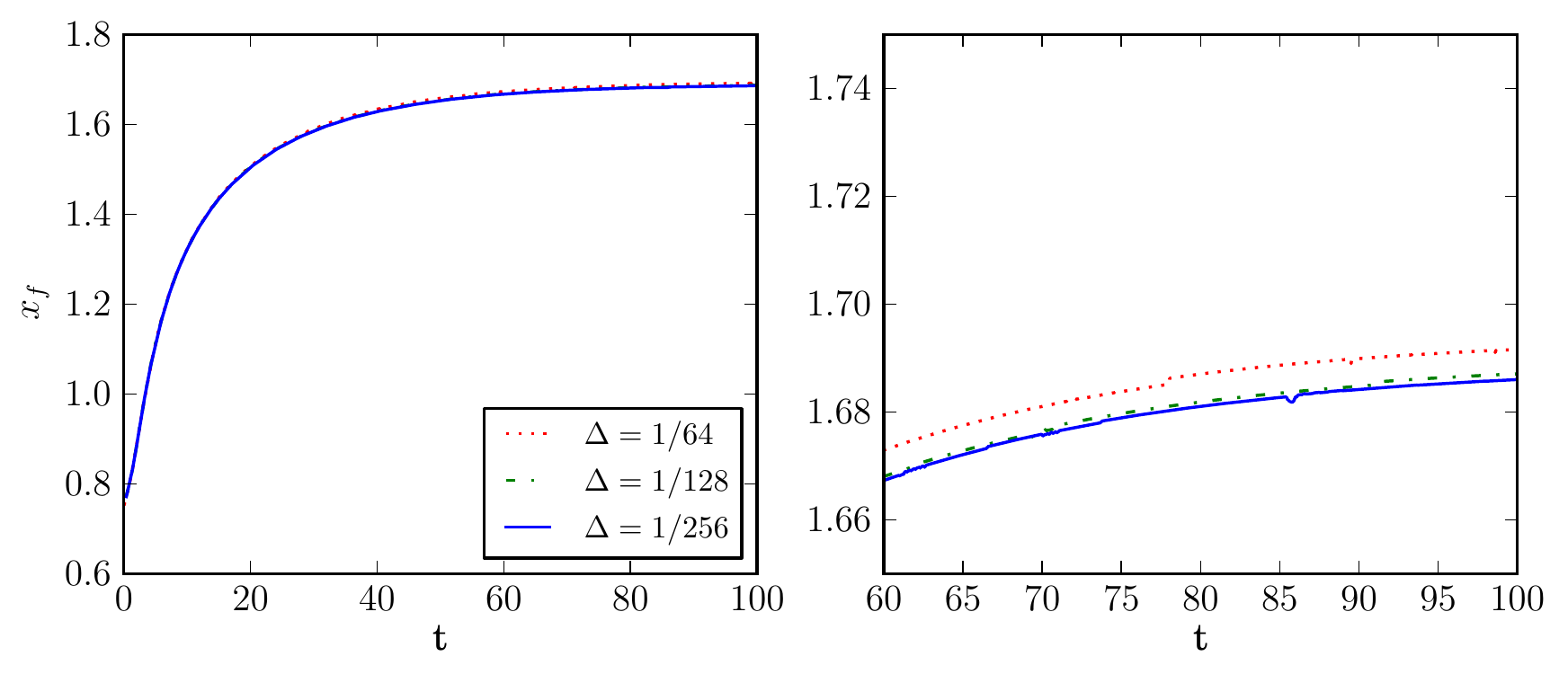}
  \caption{Method II}
 \end{subfigure}
 \caption{Front location for the spreading drop for varying resolution for (a) Methods I
   and (b) Method II.  The maximum mesh resolution is $\Delta_{max} =
   \Delta$.}
 \label{fig:front_plots}
\end{figure}

\begin{figure}
   \centering
   \includegraphics{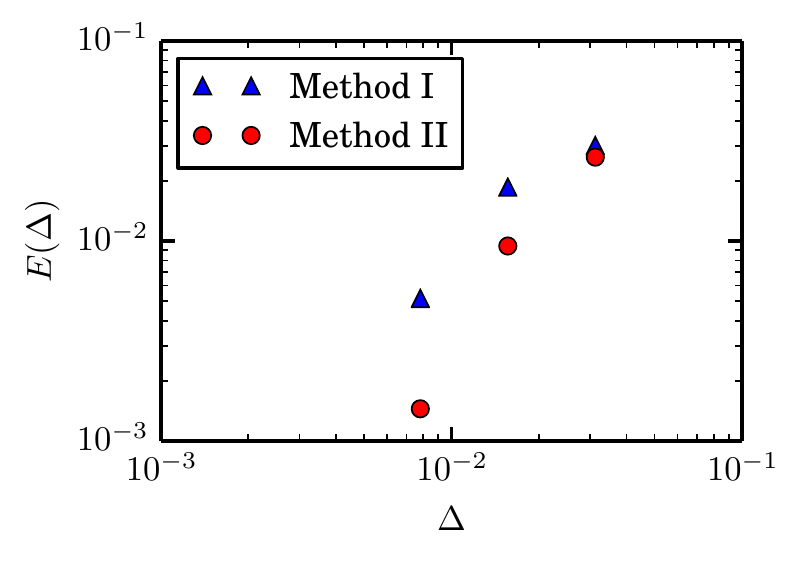}
   \caption{Convergence of the front location of spreading drops, for Methods I
     and II, where the error is estimated according to
     Eq.~\eqref{eq:front_error}.}
   \label{fig:spread_error}
\end{figure}

We now move onto the question of how a drop behaves when $\theta_i$ is far from
$\theta_{num}$.  The initial shape of the drop is as above, with $h^\ast=0.03$,
$R=0.75$, so that $\theta_i=\pi/2$.  At $y=h_c$, we use free-slip since it
allows the drop to reach its steady state more quickly, reducing computational
time.  The simulation domain is $(x,y) \in [0,4] \times [h_c,1+h_c]$.  For
these simulations, we will set $We_l=We_v=Ca_l=Ca_v=1$.  We note that even with
this choice of parameters, the dimensionless contact point speed is
sufficiently small so that surface tension effects still dominate over viscous
effects (or precisely speaking, the capillary number defined based on the speed
of the contact point is still small).  As before, we use the same material
parameters for both phases, and note that, provided that the capillary number
is small, varying the ratios between the  phases will only affect the
relaxation time; we will expand on this point below.  Here, we impose a small
contact angle of $\theta_{eq}=\pi/6$.  Unlike the above, we use an adaptive
mesh which refines regions to a level of $\Delta_{max}$ near the liquid/vapor
interface, or if there is a high gradient in $\textbf{F}_{B}$.  We will vary
$\Delta_{max}$ in what follows to study the convergence with respect to the
maximum resolution.  Figure~\ref{fig:spread_profile} shows the initial and
equilibrium profiles and pressure distribution when the drop reaches
equilibrium, computed using Method II.  The pressure inside the liquid phase is
near unity, while the pressure in the vapor phase just above the equilibrium
film is five times the capillary pressure.  The equilibrium contact angle in
Fig.~\ref{fig:spread_profile} is $\theta_{num}\approx \pi/7$.
\begin{figure}[ht]
	\centering
	\includegraphics{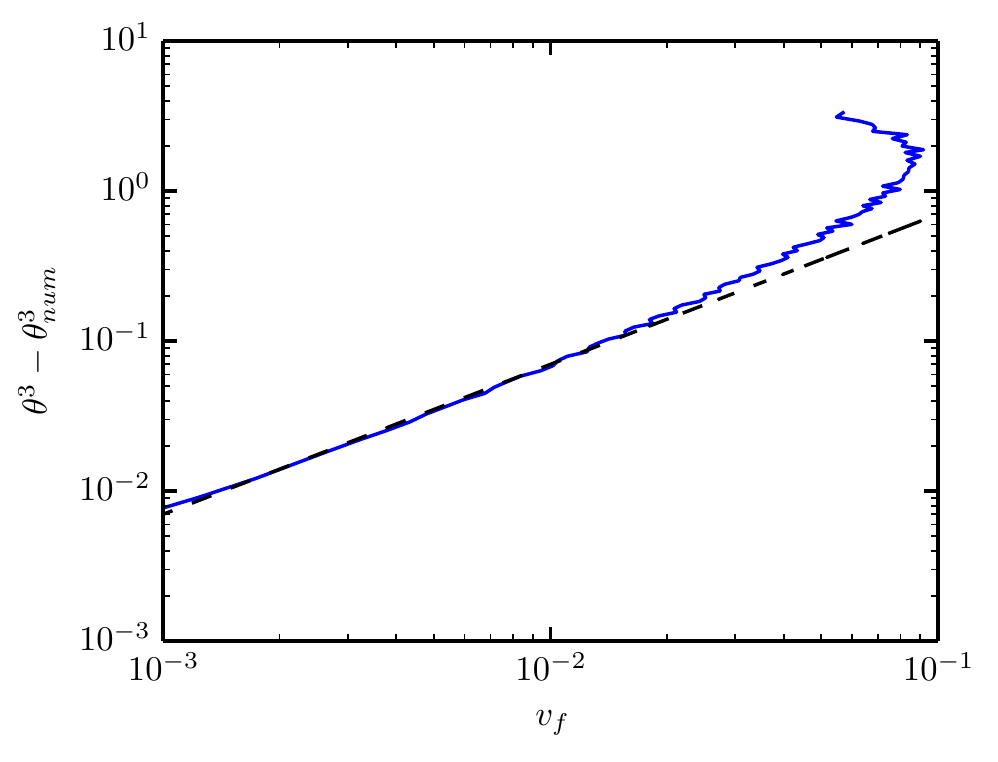}
	\caption{Cox-Voinov law for the spreading drop. The blue (solid) curve
	  shows simulation results for a droplet spreading from
	  $\theta_i=\pi/2$ to equilibrium, with an imposed contact angle of
	  $\theta_{eq}=\pi/6$ using Method II.  The black (dashed) line is
	  proportional to $v_f$, and the agreement with the blue (solid) curve
	  shows that the drop spreading approximately satisfies the Cox-Voinov
	  law.}
	\label{fig:cox_voinov}
\end{figure}

Figure~\ref{fig:front_plots} shows the front locations for the spreading drops as a
function of time for Method I and Method II.  Both methods show broadly similar results,
with Method II appearing to slightly outperform Method I in terms of convergence.  In
order to compute the convergence, we calculate the error as:
\begin{equation}\label{eq:front_error}
	E(\Delta)_{I,II} = \frac{1}{100}\int_0^{100} |{x_f(t)}_\Delta - {x_f(t)}_{\Delta=1/256} | dt
\end{equation}
Here ${x_f(t)}_\Delta$ is the front location computed with
$\Delta_{max}=\Delta$, and ${x_f(t)}_{\Delta=1/256}$ is
the front location computed with $\Delta_{max}=1/256$. 
Figure ~\ref{fig:spread_error} shows the errors.  Method II displays significantly better
convergence for the front location as a function of time.

\begin{figure}[ht]
	\centering
	\includegraphics{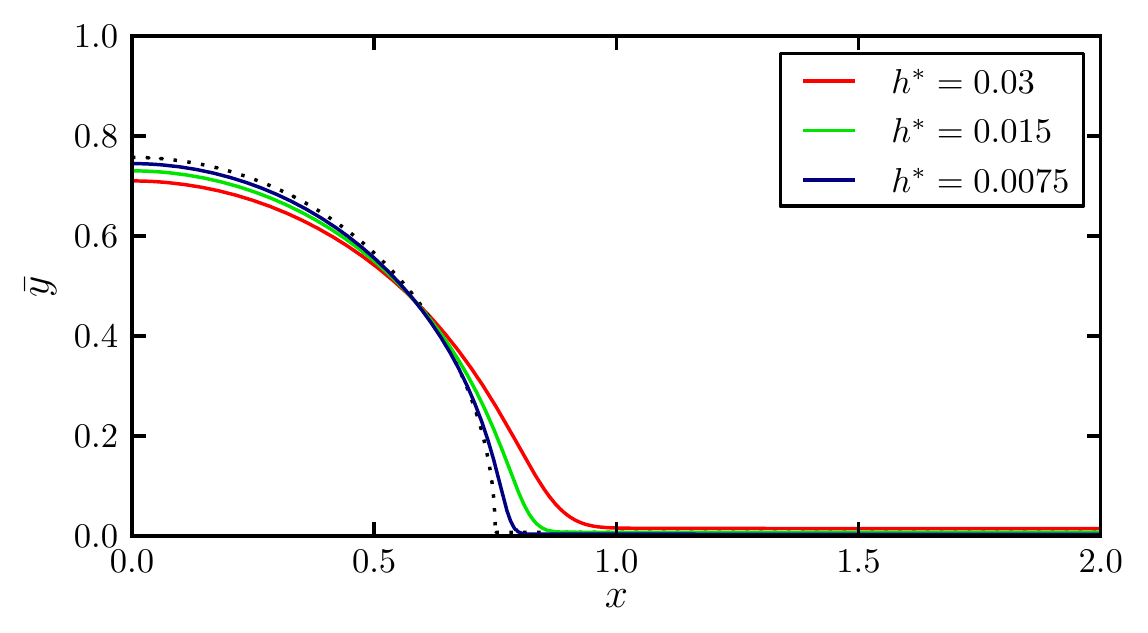}
	\caption{Equilibrium profiles with $\theta_{eq}=\theta_i=\pi/2$ for
	  various values of $h^\ast$.  The black (dotted) profile shows the
	  initial condition for $h^\ast=0.015$.  We plot $\bar{y}=y-h_c$ so that
	  for each curve the $y$ range is $(0,1)$.}
	\label{fig:conv_hstar_profiles}
\end{figure}

We briefly compare the qualitative behavior of the spreading drop to the well known
Cox-Voinov law~\cite{voinov1976}.  For a drop displacing another immiscible fluid on a
solid surface, the speed of the contact point, $v_f$, is related to $\theta_i$ and
$\theta_{num}$ to leading order by~\cite{cox1986}:
\begin{equation}\label{eq:cvlaw}
	\theta^3-\theta_{num}^3 \propto v_f
\end{equation}
Note that Eq.~\eqref{eq:cvlaw} is derived under the assumption 
that $\mu_l v_f/\sigma \ll 1$, i.e. that the capillary number defined using the front
velocity is small. Provided that one is in this regime, the choice of material parameters
only impacts the constant of proportionality in Eq.~\eqref{eq:cvlaw}.
Figure~\ref{fig:cox_voinov} shows $v_f$ versus $\theta^3-\theta_{num}^3$ using Method II,
for $\Delta_{max} = 1/256$.  The blue (solid) line shows the numerical results, and the
black (dashed) line shows the slope expected if the Cox-Voinov law is obeyed.  
We see that after initial transients $v_f$ decreases with
$\theta^3-\theta_{num}^3$, and the drop spreading approximately satisfies the  Cox-Voinov
law for $v_f \in (10^{-2},10^{-3})$.

\begin{figure}
\centering
 \begin{subfigure}{\textwidth}
  \centering
  \includegraphics{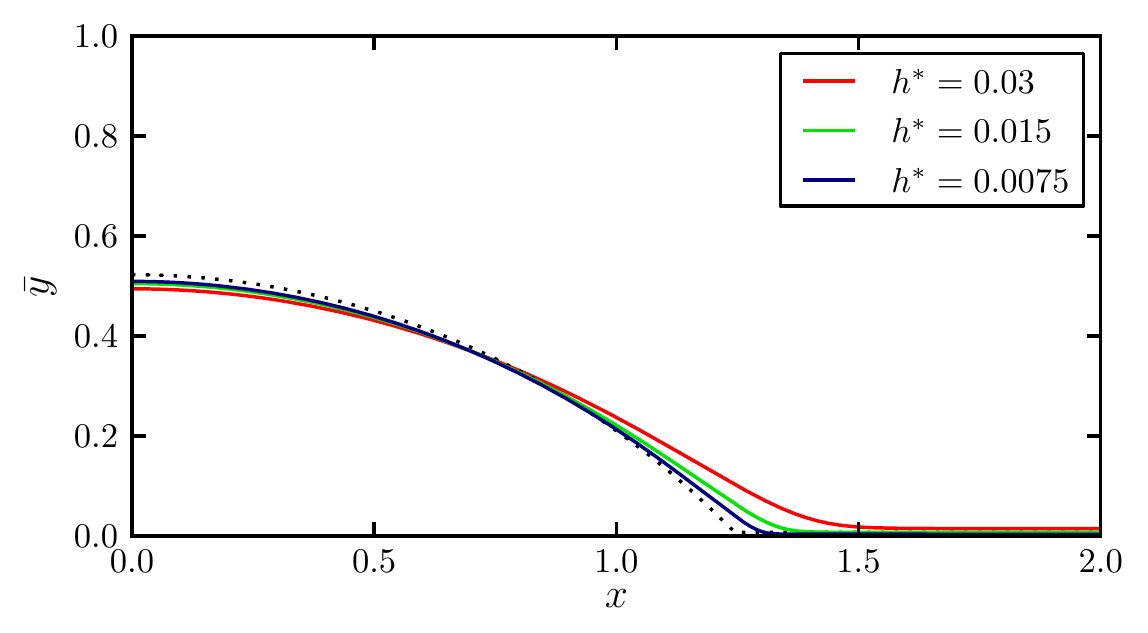}
  \caption{}
  \label{fig:dep_hstar_pi4}
 \end{subfigure}
 
 \begin{subfigure}{\textwidth}
  \centering
  \includegraphics{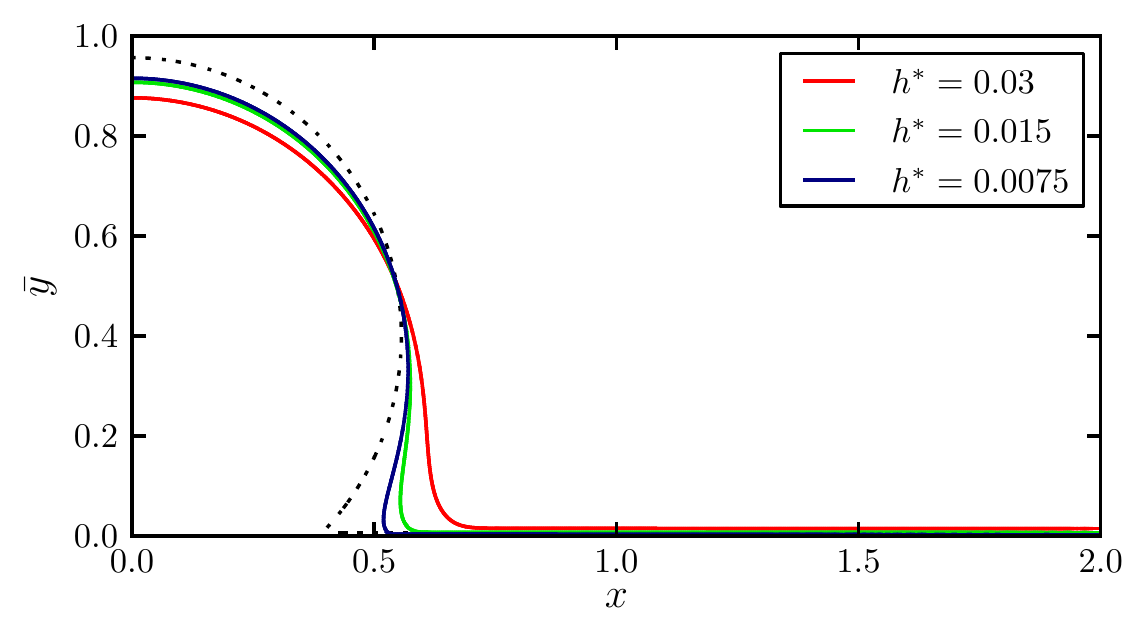}
  \caption{}
  \label{fig:dep_hstar_3pi4}
 \end{subfigure}
  \caption{Dependence of the drop profile at equilibrium on $h^\ast$ for (a)
    $\theta_{eq}=\pi/4$ and (b) $\theta_{eq}=3\pi/4$.  The black (dotted)
    profile shows the initial condition for $h^\ast=0.015$.  As in
    Fig.~\ref{fig:conv_hstar_profiles}, we plot $\bar{y}=y-h_c$ so that for
    each curve the $y$ range is $(0,1)$.}
  \label{fig:dep_hstar_theta}
\end{figure}

Next, we turn our attention to the precise value of the contact angle.  As
shown in Figs.~\ref{fig:eq_drop_profile} and \ref{fig:spread_profile}, the
actual equilibrium contact angle, $\theta_{num}$, is generally smaller than the
imposed angle $\theta_{eq}$.  This is due to the fact that
Eq.~\eqref{eq:nondim_K} is derived under the assumption of small $h^\ast$,
while in our simulations the ratio of drop radius and $h^\ast$ is about 25.  To
confirm the above statement, we analyze in more detail the resulting contact
angles as $h^*$ is varied.  We set $\theta_{eq}=\pi/2$ using Method II, and the
resolution is fixed at a uniform $\Delta=1/256$.  The value of $h^\ast$ varies
over $0.03$, $0.015$, and $0.0075$.  The initial condition is imposed with
$\theta_i=\pi/2$ as in Eq.~\eqref{eq:initial_condition}.  The drop is again
permitted to relax to its equilibrium with a fixed time step for 1.75 units of
time.  Figure~\ref{fig:conv_hstar_profiles} shows the equilibrium profiles for
various values of $h^\ast$.  The black (dotted) profile is the initial
condition for $h^\ast=0.015$, and is included as a reference.  As $h^\ast$ is
decreased, the equilibrium profiles are characterized by contact angles closer
to $\theta_{eq}$.
\begin{table}[t]
	\centering
	\begin{tabular}{l ||	l|		l|					l} 
		$h^\ast$ 	& $\theta_{num}$ & $|\theta_{num} - \theta_{eq}|/{\theta_{eq}}$ 	& $||T_i(\mathcal{C}) - T_f(\mathcal{C})||$\\
		\hline
		0.03 		&1.37  	& 0.125					& 0.05 	\\
		0.015 		&1.47   & 0.066					&0.03	\\
		0.0075		&1.53	& 0.025					&0.009	\\
	\end{tabular}
	\caption{Dependence of $\theta_{num}$ on $h^\ast$.  We calculate
	  $\theta_{num}$ using a circle fit.  The third column gives the relative
	  difference between $\theta_{num}$ and $\theta_{eq}$. The fourth column
	  is the $L^1$ norm of the difference between the initial volume
	  fractions $T_i(\mathcal{C})$ and the equilibrium $T_f(\mathcal{C})$.} 
	\label{tab:theta}
\end{table}

We quantify the dependence of $\theta_{num}$ on $h^\ast$ in
Table~\ref{tab:theta}.  As before, we compute the contact angle $\theta_{num}$
using a circle fit of the drop after $1.75$ units of time, at a fixed time step
for all simulations.  As $h^\ast$ is reduced, the calculated $\theta_{num}$
approaches $\theta_{eq}=\pi/2$.  The relative difference between $\theta_{num}$
and $\theta_{eq}$ reduces with $h^\ast$ approximately linearly.  These measures
however depend on the accuracy of the estimation of $\theta_{num}$; to analyze
the convergence more directly, we compare the volume fractions of the initial
condition, $T_i(\mathcal{C})$, with the equilibrium state $T_f(\mathcal{C})$,
using an $L^1$ norm computed as Eq.~\eqref{eq:l1norm}.  Initially,
$\theta_i=\theta_{eq}$, so this comparison provides a measure of the difference
between $\theta_{num}$ and $\theta_{eq}$.  This difference again decreases
approximately linearly with $h^*$.

Finally, we consider the effects of varying $h^\ast$ for $\theta_{eq}$ other
than $\pi/2$.  Figure \ref{fig:dep_hstar_pi4} shows the equilibrium drop
profiles for $\theta_{eq}=\theta_i=\pi/4$ and varying $h^\ast$, and
Fig.~\ref{fig:dep_hstar_3pi4} shows the same for $\theta_{eq}=\theta_i=3\pi/4$.
For $\theta_{eq}=\pi/4$, as $h^\ast$ decreases, the profile quickly converges
to the initial condition and hence the contact angle to $\theta_{eq}$, as seen
by the comparison of the black (dotted) curve with that of $h^\ast=0.0075$ in
Fig.~\ref{fig:dep_hstar_pi4}.  For $\theta_{eq}=3\pi/4$, it can be seen that
even for $h^\ast=0.0075$, the drop profile still shows some difference between
$\theta_{num}$ and $\theta_{eq}$.  Nonetheless, this plot demonstrates a
central advantage of our methods: $\theta_{num}$ larger than $\pi/2$ can be
simulated with the van der Waals force. 

\section{Conclusions}\label{sec:conc}

In this paper, we have described a novel approach for including the fluid/solid
interaction forces, into a direct solver of the Navier-Stokes equations with a
volume of fluid interface tracking method. The model does not restrict the
contact angles to be small, and therefore, can be used to accurately model
wetting and dewetting of fluids on substrates characterized by arbitrary
contact angles. We study the problem of a two-dimensional drop on a substrate
and compare the results with the Cox-Voinov law for drop spreading. These
validations and results illustrate the applicability of our proposed method to
model flow problems involving contact lines. Furthermore, our approach has the
desirable property of regularizing the viscous stress singularity at a moving
contact line since it naturally introduces an equilibrium fluid film.  

We have considered two alternative finite-volume discretizations of the van der
Waals force term that enters the governing equations to include the fluid/solid
interaction forces. These two methods are complementary in terms of their
accuracy and the ease of use, and therefore the choice of the method could be
governed by the desired features of the results. In particular, Method I
discretizations can be straightforwardly extended to the third spatial
dimension. However, we have shown that, when implementing Method I, to
guarantee accurate results, sufficient spatial resolution of the computational
mesh must be used.  We also show that Method II does not suffer accuracy
deterioration at low mesh resolutions, and is therefore, superior to Method I,
and furthermore outperforms Method I for spreading drops at all resolutions,
albeit being more involved to implement. 

The presented approach opens the door for modeling problems that could not be
described so far, in particular dewetting and associated film breakup for
fluids characterized by large contact angles. Furthermore, the model allows to
study the effects of additional mechanisms, such as inertia.  We will consider
these problems in future work.

\section*{Acknowledgements}

This research was supported by the The National Science Foundation
under grants NSF-DMS-1320037 and CBET-1235710.
The authors acknowledge many useful discussions with Javier Diez of Universidad Nacional del Centro de la
Provincia de Buenos Aires, Argentina.

\end{document}